\documentstyle[sprocl]{article}

\begin{document}
\title{CONSTRUCTION OF LOW-ENERGY EFFECTIVE ACTION IN ${\cal N}$=4 SUPER
YANG-MILLS THEORIES}

\author{I.L. Buchbinder}

\address{Department of Theoretical Physics Tomsk State Pedagogical
University \\ Tomsk, 634041, Russia}

\maketitle

\begin{abstract}
We review a recent progress in constructing
low-energy effective action in ${\cal N}$=4 super Yang-Mills theories.
Using harmonic superspace approach we consider ${\cal N}$=4 SYM in terms
of unconstrained ${\cal N}$=2 superfield and apply ${\cal N}$=2 background field
method to finding effective action for ${\cal N}$=4 SU(N) SYM broken down
to U(1)$^{N-1}$.  General structure of leading low-energy corrections
to effective action is discussed.
\end{abstract}

\section{Introduction}

Low-energy structure of quantum supersymmetric field theories is
described by the effective lagrangians of two types: chiral and general
or holomorphic and non-holomorphic.  Non-holomorphic or general
contributions to effective action are given by integrals over full
superspace while holomorphic or chiral contributions are given by
integrals over chiral subspace of superspace. As a result the effective
action in low-energy limit is defined by the chiral superfield ${\cal
F}$ which is called holomorphic or chiral effective potential and real
superfield ${\cal H}$ which is called non-holomorphic or general
effective potential.

Possibility of holomorphic corrections to effective
action was firstly demonstrated in [1] ( see also [2]) for ${\cal N}$=1
SUSY and in [3] for ${\cal N}$=2 SUSY.  The modern interest to structure
of low-energy effective action in extended supersymmetric theories was
inspired by the seminal papers [4] where exact instanton contribution
to holomorphic effective potential has been found for ${\cal N}$=2 SU(2)
super Yang-Mills theory.  These results have later been extended for
various gauge groups and for coupling to matter (see e.g. [5]).  One
can show that in generic ${\cal N}$=2 SUSY models namely the holomorphic
effective potential is leading low-energy contribution. Non-holomorphic
potential is next to leading correction. A detailed investigation of
structure of low-energy effective action for various ${\cal N}$=2 SUSY
theories has been undertaken in [6-9].

A further study of quantum aspects of
supersymmetric field models leads to problem of effective action in
${\cal N}$=4 SUSY theories. These theories being maximally extended
global supersymmetric models possess the remarkable properties on
quantum level: (i) ${\cal N}$=4 super Yang-Mills model is finite quantum
field theory, (ii) ${\cal N}$=4 super Yang-Mills model is superconformal
invariant theory and hence, its effective action can not depend on any
scale.  These properties allow to analyze a general form of low-energy
effective action and see that it changes drastically in compare with
generic ${\cal N}$=2 super Yang-Mills theories.

Analysis of structure of low-energy effective action in ${\cal N}$=4
SU(2) SYM model spontaneously broken down to U(1) has been fulfilled in
recent paper by Dine and Seiberg [10]. They have investigated a part of
effective action depending on ${\cal N}$=2 superfield strengths $W$,
$\bar{W}$ and shown
(i)
Holomorphic quantum corrections are trivial in ${\cal N}$=4 SYM.
Therefore, namely non-holomorphic effective potential is leading
low-energy contribution to effective action, (ii)
Non-holomorphic effective potential ${\cal H}(W,\bar W)$ can be found
on the base of the properties of quantum ${\cal N}$=4 SYM theory up to a
coefficient.  All perturbative or non-perturbative corrections do not
influence on functional form of ${\cal H}(W,\bar W)$ and concern only
of this coefficient.

The approaches to direct calculation of non-holomorphic effective
potential including the above coefficient have been developed in
[11-13], extensions for gauge group SU(N) spontaneously broken to
maximal torus have been given in [15-17] (see also [14] where some
bosonic contributions to low-energy effective action have been found).

\section{${\cal N}$=4 super Yang-Mills theory in harmonic superspace}

As well known, the most powerful and adequate approach to investigate
the quantum aspects of supersymmetric field theories is formulation of
these theories in terms of unconstrained superfields carrying out a
representation of the supersymmetry. Unfortunately such a manifestly
${\cal N}$=4 supersymmetric formulation for ${\cal N}$=4 Yang-Mills theory
is still unknown.  A purpose of this paper is study a structure of
low-energy effective action for ${\cal N}$=4 SYM as a functional of ${\cal N}$=2
superfield strengths. In this case it is sufficient to realize the
${\cal N}$=4 SYM theory as a theory of ${\cal N}$=2 unconstrained
superfields. It is naturally achieved within harmonic superspace. The
${\cal N}$=2 harmonic superspace [19] is the only manifestly ${\cal N}$=2
supersymmetric formalism allowing to describe general ${\cal N}$=2
supersymmetric field theories in terms of unconstrained ${\cal N}$=2
superfields.  This approach has been successfully applied to problem of
effective action in various ${\cal N}$=2 models in recent works [7, 9,
12, 13, 16, 18, 20].

From point of view of ${\cal N}$=2 SUSY, the ${\cal N}$=4
Yang-Mills theory describes interaction of ${\cal N}$=2 vector multiplet
with hypermultiplet in adjoint representation. Within harmonic
superspace approach, the vector multiplet is realized by unconstrained
analytic gauge superfield $V^{++}$. As to hypermultiplet, it can be
described either by a real unconstrained superfield $\omega$
($\omega$-hypermultiplet) or by a complex unconstrained analytic
superfield $q^+$ and its conjugate ($q$-hypermultiplet). In the
$\omega$-hypermultiplet realization, the classical action of ${\cal N}$=4
SYM model has the form

$$
S[V^{++},\omega]=\frac{1}{2g^2}{\rm tr}\int d^4xd^4\theta
W^2- \frac{1}{2g^2}{\rm tr}\int
d\zeta^{(-4)}\nabla^{++}\omega \nabla^{++}\omega \eqno(1)%2.1
$$
The first terms here is pure ${\cal N}$=2 SYM action and the second term
is action $\omega$-hypermultiplet. In $q$-hypermultiplet realization,
the action of the ${\cal N}$=4 SYM model looks like this

$$
S[V^{++},q^+,\stackrel{\smile}{q}^+]= \frac{1}{2g^2}{\rm
tr}\int d^4xd^4\theta W^2- \frac{1}{2g^2}{\rm tr}\int
d\zeta^{(-4)}q^{+i}\nabla^{++}q^+_i \eqno(2)%2.2
$$
where
$$
q^+_i=(q^+,\stackrel{\smile}{q}^+),\qquad
q^{i+}=\varepsilon^{ij}q^+_j=(\stackrel{\smile}{q}^+,-q^+)
\eqno(3)%2.3
$$
All other denotions are given in [19].  Both models (1,2) are
equivalent and manifestly ${\cal N}$=2 supersymmetric by
construction. However, as has been shown in [19], both these models
possess hidden ${\cal N}$=2 supersymmetry and as a result they actually
are ${\cal N}$=4 supersymmetric.

\section{General form of non-holomorphic effective potential}

We study the effective action $\Gamma$ for ${\cal N}$=4 SYM with gauge
group SU(2) spontaneously broken down to U(1). This effective action is
considered as a functional of ${\cal N}$=2 superfield strengths $W$ and
$\bar W$. Then holomorphic effective potential ${\cal F}$ depends on
chiral superfield $W$ and it is integrated over chiral subspace of ${\cal N}$=2
superspace with the measure $d^4x\,d^4\theta$. Non-holomorphic
effective potential ${\cal H}$ depends on both $W$ and $\bar W$. It is
integrated over full ${\cal N}$=2 superspace with the measure $d^4x\,d^8\theta$.
Taking into account the mass dimensions of $W$, ${\cal F}(W)$, ${\cal
H}(W,\bar W)$ and the superspace measures $d^4x\,d^4\theta$ and
$d^4x\,d^8\theta$ ones write

$$
{\cal F}(W)=W^2f\left(\frac{W}{\Lambda}\right),\qquad {\cal H}(W,\bar
W)={\cal H}\left(\frac{W}{\Lambda}, \frac{\bar W}{\Lambda}\right)
\eqno(4)%3.1
$$
where $\Lambda$ is some scale and $f(\frac{W}{\Lambda})$ and ${\cal
H}(\frac{W}{\Lambda},\frac{\bar W}{\Lambda})$ are the dimensionless
functions of their arguments. The effective action is scale
independent, therefore

$$
\Lambda\frac{d}{d\Lambda} \int
d^4x\,d^4\theta W^2f\left(\frac{W}{\Lambda}\right)=0, %3.2
\quad
\Lambda\frac{d}{d\Lambda} \int d^4x\,d^8\theta{\cal
H}\left(\frac{W}{\Lambda}, \frac{\bar W}{\Lambda}\right)=0
\eqno(5)%3.3
$$
First of eqs (5) leads to $f(\frac{W}{\Lambda})=const$. Second of eqs
(5) reads

$$
\Lambda\frac{d}{d\Lambda}{\cal H}=g\left(\frac{W}{\Lambda}\right)+ \bar
g\left(\frac{\bar W}{\Lambda}\right)\eqno (6)%3.4
$$
Here $g$ is arbitrary chiral function of chiral superfield
$\frac{W}{\Lambda}$ and $\bar g$ is conjugate function.
Since $f(\frac{W}{\Lambda})=const$ the
holomorphic effective potential ${\cal F}(W)$ is
proportional to classical lagrangian $W^2$. General solution to eq (6)
is written as follows
$$
{\cal
H}\left(\frac{W}{\Lambda},\frac{\bar W}{\Lambda}\right)=
c\log\frac{W^2}{\Lambda^2}\log\frac{\bar W^2}{\Lambda^2}\eqno(7)%3.4
$$
with arbitrary coefficient $c$. As a result, holomorphic effective
potential is trivial in ${\cal N}$=4 SYM theory. Therefore, namely
non-holomorphic effective potential is leading low-energy quantum
contribution to effective action.  Moreover, the non-holomorphic
effective potential is found exactly up to coefficient and given by eq
(7) [10]. Any perturbative or non-perturbative quantum corrections are
included into a single constant $c$.  However, this result immediately
face the problems:
1) is there exist a calculational procedure
allowing to derive ${\cal H}(W/\Lambda,\bar W/\Lambda)$ in form (7)
within a model?
2) what is value of $c$? If $c=0$, the non-holomorphic
effective potential vanishes and low-energy effective action in ${\cal N}$=4 SYM
is defined by the terms in effective action depending on the covariant
derivatives of $W$ and $\bar{W}$,
3) what is a structure of
non-holomorphic effective potential for the other then SU(2) gauge
groups?

The answers all these questions have been given in [11-17]. Further we
are going to discuss a general manifestly ${\cal N}$=2 supersymmetric and gauge
invariant procedure of deriving the non-holomorphic effective potential
in one-loop approximation [13,16]. This procedure is based on the
following points:
1) formulation of ${\cal N}$=4 SYM theory in terms of ${\cal N}$=2
unconstrained superfields in harmonic superspace [19],
2) ${\cal N}$=2
background field method [9] providing manifest gauge invariance on all
steps of calculations,
3) identical transformation of path integral for
effective action over ${\cal N}$=2 superfields to path integral over some
${\cal N}$=1 superfields. This point is nothing more then replacement of
variables in path integral, 4) superfield proper-time technique
(see first of refs [2]) which is
manifestly covariant method for evaluating effective action in
superfield theories.

\section{Non-holomorphic effective potential for SU(N)-gauge group}

We study effective action for the classically equivalent theories (1,
2) within ${\cal N}$=2 background field method [9]. We assume also that
the gauge group of these theories is SU(N). In accordance with
background field method [9], the one-loop effective action in both
realizations of ${\cal N}$=4 SYM is given by

$$
\Gamma^{(1)}[V^{++}]=\frac{i}{2
}{\rm Tr}_{(2,2)} \log\stackrel{\frown}{\Box}-\frac{i}{2}{\rm
Tr}_{(4,0)} \log\stackrel{\frown}{\Box} \eqno(8)%4.1
$$
where
$\stackrel{\frown}{\Box}$ is the analytic d'Alambertian introduced in
[9].
$$
\begin{array}{rcl} \stackrel{\frown}{\Box}&=&{\cal
D}^m{\cal D}_m+ \displaystyle\frac{i}{2}({\cal D}^{+\alpha}W){\cal
D}^-_\alpha+ \displaystyle\frac{i}{2}(\bar{\cal D}^+_{\dot{\alpha}}\bar
W){\bar {\cal D}}^{-\dot{\alpha}}-\\ &-&\displaystyle\frac{i}{4}({\cal
D}^{+\alpha} {\cal D}^+_\alpha W){\cal
D}^{--}+ \displaystyle\frac{i}{8}[{\cal
D}^{+\alpha},{\cal D^-_\alpha}]W+\displaystyle\frac{i}{2}
\{\bar{W},W\} \end{array}\eqno(9)%4.2
$$
The formal definitions of the ${\rm
Tr}_{(2,2)}\log\stackrel{\frown}{\Box}$ and ${\rm
Tr}_{(4,0)}\log\stackrel{\frown}{\Box}$ are given in [12]. Our purpose
is finding of non-holomorphic effective potential ${\cal H}(W,\bar W)$
where the constant superfields $W$ and $\bar W$ belong to Cartan
subalgebra of the gauge group SU(N). Therefore, for calculation of
${\cal H}(W,\bar W)$ it is sufficient to consider on-shell background,
${\cal D}^{\alpha(i}{\cal D}^{j)}_\alpha W=0$.
In this case the one-loop effective action (8) can be written in the
form [16]

$$
\Gamma^{(1)}=\sum\limits_{k<l}\Gamma_{kl},\qquad \Gamma_{kl}=i{\rm
Tr}\log\Delta_{kl} \eqno(10)%4.6
$$
$$
\Delta_{kl}={\cal D}^m{\cal
D}_m-(W^{k\alpha}-W^{l\alpha}) {\cal
D}_\alpha+(\bar{W}^k_{\dot{\alpha}}-\bar{W}^l_{\dot{\alpha}}) \bar{\cal
D}^{\dot{\alpha}}+|\Phi^k-\Phi^l|^2 \eqno(11)%4.7)
$$
and ${\cal D}_m$,
${\cal D}_\alpha$, $\bar{\cal D}_{\dot\alpha}$ are the ${\cal N}$=1
supercovariant derivatives. Here

$$
\Phi={\rm diag}(\Phi^1,\Phi^2,\dots,\Phi^N),\qquad
\sum\limits_{k=1}^N \Phi^k=0. \eqno(12)%4.8
$$
$$ W_\alpha={\rm
diag}(W_\alpha^1,\dots,W_\alpha^N),\qquad \sum\limits_{k=1}^N
W_\alpha^k = 0
$$
$\Phi$ and $W_\alpha$ are the ${\cal N}$=1 projections of $W$.
The operator (11) has been introduced in [16].
Evaluation of the ${\rm Tr}\log\Delta_{kl}$ leads to
$$
\Gamma_{kl}=\frac{1}{(4\pi)^2}\int d^8z\frac{W^{\alpha kl}W^{kl}_\alpha
\bar{W}^{kl}_{\dot{\alpha}}\bar{W}^{\dot{\alpha}kl}}
{(\Phi^{kl})^2(\bar{\Phi}^{kl})^2} \eqno(13)%4.9
$$
where
$$
\Phi^{kl}=\Phi^k-\Phi^l,\qquad W^{kl}=W^k-W^l \eqno(14)%4.10
$$
Eqs (10,
13, 14) define the non-holomorphic effective potential of ${\cal N}$=4 SYM
theory in terms of ${\cal N}$=1 projections of ${\cal N}$=2 superfield
strengths.  The last step is restoration of ${\cal N}$=2 form of
effective action (13). For this purpose we write contribution of
non-holomorphic effective potential to effective action in terms of
covariantly constant ${\cal N}$=1 projections $\Phi$ and $W_\alpha$

$$
\int d^{12}z{\cal
H}(W,\bar{W})=\int d^8zW^\alpha W_\alpha
\bar{W}_{\dot{\alpha}}\bar{W}^{\dot{\alpha}}\frac{\partial^4{\cal
H}(\bar{\Phi},\Phi)}{\partial\Phi^2\partial\bar{\Phi}^2}+ {\rm
derivatives} \eqno(15)%4.11
$$
Comparison of eqs (14) and (15) leads to
$$
\Gamma^{(1)}=\int d^4xd^8\theta{\cal H}(\bar{W},W)
$$
$$
{\cal H}(W,\bar{W})=\frac{1}{(8\pi)^2}\sum\limits_{k<l}
\log\left(\frac{(\bar{W}^k-\bar{W}^l)^2}{\Lambda^2}\right)
\log\left(\frac{(W^k-W^l)^2}{\Lambda^2}\right) \eqno(16)%4.12
$$
Eq (16) is our final result. In partial case of SU(2) group
spontaneously broken down to U(1) eq (16) coincides with eq (7) where
$c=1/(8\pi^2)$. Another approach to derivating the effective potential
(16) for SU(2)-group was developed in recent paper [13].

\section{Discussion}

Eq (16) defines the non-holomorphic effective potential depending on
${\cal N}$=2 superfield strengths for ${\cal N}$=4 SU(N) super Yang-Mills
theories. As a result we answered all the questions formulated in
section 3. First, we have presented the calculational procedure
allowing to find non-holomorphic effective potential. Second, we
calculated the coefficient c in eq (7) for SU(2) group. It is equal to
$1/(8\pi)^2$.  Third, a structure of non-holomorphic effective
potential for the gauge group SU(N) has been established.

It is
interesting to point out that the scale $\Lambda$ is absent when the
non-holomorphic effective potential (16) is written in terms of ${\cal
N}=1$ projections of $W$ and $\bar W$ (see eqs (15, 16)). Therefore, the
$\Lambda$ will be also absent if we write the non-holomorphic effective
potential through the component fields. We need in $\Lambda$ only to
present the final result in manifestly ${\cal N}$=2 supersymmetric form.
${\cal N}$=1 form of non-holomorphic effective potential (10) allows very
easy to get leading bosonic component contribution. Schematically it
has the form $F^4/|\phi|^4$, where $F_{mn}$ is abelian strength
constructed from vector component and $\phi$ is a scalar component of
${\cal N}$=2 vector multiplet. It means that non-zero expectation value of
scalar field $\phi$ plays a role of effective infrared regulator in
${\cal N}$=4 SYM theories.

Generalization of low-energy effective action
containing all powers of constant $F_{mn}$ has recently
been constructed in [18]. The
direct proof of absence of three- and four-loop corrections to ${\cal
H}$ was given in [20].

It was recently noticed [17, 21] that R-symmetry and scale independence
do not prohibit the terms of the form $G(W^{ij}/W^{kl},\bar W^{ij}/\bar
W^{kl})$ in non-holomorphic effective potential for SU(N)-models with
$N>2$
where G is a real function and $W^{ij}=W^{i}-W^{j}$. However, the
calculations [20] did not confirm an emergence of such terms at one-,
two-, three- and four loops. One can expect that ${\cal N}$=4
supersymmetry imposes more rigid restrictions on a structure of
effective potential then only R-symmetry and scale independence
taking place for any ${\cal N}$=2 superconformal theory.

\noindent{\bf Acknowledgments.}
I am very grateful to E.I. Buchbinder, E.I. Ivanov, S.M. Kuzenko, B.A.
Ovrut, A.Yu Petrov, A.A. Tseytlin for collaboration. The work was
supported in part by the RFBR grant 99-02-16617, RFBR-DFG grant
99-02-04022, INTAS grant 991-590 and 001-254.

\end{document}